\documentclass[useAMS,usenatbib]{mn2e}
\usepackage{times}
\usepackage{epsfig}
\usepackage{amssymb}

\def\be{\begin{equation}}
\def\ee{\end{equation}}
\def\bc{\begin{center}}
\def\ec{\end{center}}
\def\beq{\begin{eqnarray}}
\def\eeq{\end{eqnarray}}

\def\RL{R_{\rm L}}
\def\Bj{B_{\rm j}}
\def\Rj{R_{\rm j}}
\def\Be{B_{\rm e}}
\def\gammac{\gamma_{\rm c}}
\def\gmaxc{\gamma_{\rm c, \max}}

\def\phe{\epsilon}
\def\phem{\epsilon_{\max}}
\def\phx{x}

\def\phei{\epsilon_{\rm e}}
\def\phel{\epsilon_{\rm j}}
\def\phelmin{\epsilon_{\rm j, \min}}
\def\pheimin{\epsilon_{\rm e, \min}}

\def\xmin{\phx_{\min}}
\def\xmax{\phx_{\max}}

\def\aj{a_{\rm j}}
\def\ai{a_{\rm e}}

\def\absiso{a^{\rm iso}_{\gamma\gamma}}
\def\absgg{a_{\gamma\gamma}}
\def\absggmax{a_{\gamma\gamma,\max}}
\def\taugg{\tau_{\gamma\gamma}}
 \def\xkn{\phx_{\rm KN}} 
 \def\xlow{\phx_{\rm low}}

\def\phic{\phi_{\rm c}}
\def\betac{\beta_{\rm c}}
\def\sigmat{\sigma_{\rm T}}
\def\taut{\tau_{\rm T}}

\def\d{{\rm d}}
\def\rme{{\rm e}}
\def\alphaiso{\alpha}

\def\etaiso{\eta_{\rm i}} 
\def\etab{\eta_{\rm B}} 
\def\etaj{\eta_{\rm K}} 

\def\Ljet{L_{\rm K}}
\def\Ldisc{L_{\rm d}}
\def\Ld{L_{\rm d,45}}
\def\Lb{L_{\rm B}}

\def\UB{U_{\rm B}}
\def\Urad{U_{\rm rad}}
\def\UT{U_{\rm T}}
\def\Uc{U_{\rm cycle}}
\def\RL{R_{\rm L}}
\def\Rin{R_{\rm in}}
\def\Rout{R_{\rm out}}

\def\Fd{F_{\rm d}}
\def\Fiso{F_{\rm iso}}
\def\mp{m_{\rm p}}
\def\me{m_{\rm e}}
\def\np{n_{\rm p}}

\title[Photon breeding  in relativistic astrophysical jets]
{A photon breeding mechanism for the high-energy emission of relativistic jets}

\author[Boris E. Stern and Juri Poutanen]
{Boris~E.~Stern$^{1,2,3}$\thanks{E-mail:
stern@bes.asc.rssi.ru (BES), juri.poutanen@oulu.fi (JP)}
and Juri~Poutanen$^{3}$\footnotemark[1] \\
$^{1}$Institute for Nuclear Research, Russian Academy of Sciences,
Prospect 60-letija Oktjabrja 7a, Moscow 117312, Russia\\
$^{2}$Astro Space Center, Lebedev Physical Institute,
Profsoyuznaya 84/32,  Moscow 117997, Russia\\
$^{3}$Astronomy Division, P.O.Box 3000, FIN-90014 University of Oulu,
Finland}

\begin{document}
\date{Accepted, Received}
\pagerange{\pageref{firstpage}--\pageref{lastpage}} \pubyear{2005}
\maketitle

\label{firstpage}

\begin{abstract}
We propose a straightforward and efficient mechanism for the high-energy
emission of relativistic astrophysical jets associated with an exchange of interacting
high-energy photons between the jet and the external environment. Physical processes
 playing the main role in this mechanism are electron-positron pair production by
photons and the inverse Compton scattering. This scenario has been studied analytically
as well as with numerical simulations demonstrating that a relativistic jet (with the Lorentz 
factor larger than 3--4) moving through 
the sufficiently dense, soft radiation field inevitably undergoes transformation into a luminous state. 
The process has a supercritical character: the high-energy photons breed exponentially being fed
directly by the bulk kinetic energy of the jet. Eventually particles feed back
on the fluid dynamics and the jet partially decelerates.
As a result, a significant fraction (at least 20 per cent)
of the jet kinetic energy is converted into radiation mainly in the MeV -- GeV energy range. 
The mechanism maybe responsible for the bulk of the emission of relativistic jets 
in active galactic nuclei, microquasars and gamma-ray bursts.
\end{abstract}

%------------------------------------
%\keywords

%-----------------------------------

\section{Introduction}

Traditionally, the dissipation of a relativistic bulk motion into radiation
is assumed to be associated with the shock acceleration of {\it charged}
particles.  In a relativistic case its role is limited by a number of factors
\citep[see e.g.][]{a01,bo99}.   \citet{d03} and \citet{s03} independently suggested
that interacting {\it neutral} particles can convert bulk kinetic energy
into radiation much more efficiently than this can be done by charge particles.
Indeed, the neutral particles easily cross the shock front or the boundary of the shear layer in both directions and can be converted into charge particles for example via $e^\pm$ pair
productions by two photons.
Pairs in turn convert their energy into photons by Compton scattering,
some of which cross the boundary again. Such exchange of particles between
media moving respectively to each other with high Lorentz factor works in the 
same way  as  Fermi acceleration, but more efficiently as 
there is no problem with diffusion through the boundary layer. 
This process can proceed in a runaway manner where the high-energy 
photons breed exponentially similarly to neutron breeding in nuclear chain reaction.  

\citet{s03} demonstrated with numerical simulations that in the case of an
ultra-relativistic shock in a moderately dense medium (above $\sim 10^4$
particles cm$^{-3}$) this process  leads to a dramatic increase in the shock 
high-energy emission (electromagnetic catastrophe) and eventually to the
elimination of the shock front, which is converted into a smooth radiation front.
Such effect can take place in gamma-ray bursts.
In this work we consider the same scenario for a different kind of
relativistic fluid: a shear flow  in astrophysical jets.
While we specifically consider here jets in active galactic nuclei (AGNs) 
which emit tremendous power in form of the
gamma-rays in the MeV--TeV energy range, 
the mechanism can operate in  the jets from microquasars as well as 
in gamma-ray bursts.

There exist different models of the jet emission mechanism,
probably the most popular one is associated with internal
shocks in the jet \citep{r78,px94,rm94}. 
Nevertheless, the maximal variation of $\Gamma$
is at the jet boundary and it is there we can expect the
most intensive energy release.
The jet bulk energy can be dissipated in principle through the acceleration 
of  charged particles \citep{b90,o00,so02}. 
If Thomson opacity is  sufficiently large, the standard radiation viscosity can also cause 
the energy dissipation in the boundary layer as, for example, discussed  by \citet{ab92}. 
In the case of AGN jets the Thomson opacity is, however, insufficient to
provide the efficient reflection of photons crossing the boundary of the jet. 
Here we show  that the role of the ``mirror'' can be played by the soft radiation, 
which provides the pair production opacity for the high-energy photons. 

We study the mechanism of shear flow energy dissipation
numerically in the same way as it was done by \citet{s03} for shocks.
The problem formulation
differs in its geometry (tangential jump of the fluid velocity instead of
the head-on shock) and in the character of the primary photon background.
The complete solution of the problem should account for the feedback
of particles on the fluid dynamics and requires detailed numerical treatment
of the hydrodynamical part of the problem.
In this first study we neglect internal pressure and consider a 
one-dimensional dust approximation.
With this simplified treatment of hydrodynamics  we formulate our problem 
in the following way. We start from a non-radiating, idealized cylindrical, 
homogeneous jet and then track  its evolution to check whether it
produces a runaway photon breeding or not. 
This numerical experiment is sufficient to demonstrate
that under certain circumstances a non-radiating jet cannot exist:
it must radiate away a substantial fraction of its kinetic energy.

 We present the model of  the jet  and the environment and formulate the quantitative 
problem to be solved in Section \ref{sec:model}.
 In Section \ref{sec:anal}, we describe the proposed mechanism qualitatively, 
formulate the necessary conditions needed for its operation, and 
provide a simplified analytical study of separate parts of the mechanism. 
 In Section \ref{sec:numer}, we give the details of the numerical method for simulation 
of the entire process. 
 The results of numerical simulations for two representative sets of parameters, 
 which   show a runaway regime are presented in Section \ref{sec:simul}.  We discuss 
 the results, problems with the model and possible effects in more realistic 
 situations in Section \ref{sec:disc}, and we conclude in Section \ref{sec:concl}.

\section{Model of the jet and the environment}
\label{sec:model}

We consider a  jet of Lorentz factor $\Gamma$ 
consisting of cold electrons and protons and the magnetic field.
We consider a piece of the jet  centered at distance $R$ from the central 
source and length $20 \Rj$, where the jet radius  is $\Rj=R \theta$ and 
$\theta$ has the meaning of the opening angle, while 
for simplicity we approximate the jet  by a cylinder of radius $\Rj$.
We choose parameters corresponding to the case of AGN
taking the distance scale $R \sim 10^{17}$ cm,  
$\Gamma\sim 10$--$20$ and  $\theta=0.05$. 
The radial distance from the jet axis $r$ and  the distance from the central source $z$ 
is measured in units of  $\Rj$.  The  unit of time is $\Rj/c$.

The jet propagates through the soft  radiation field, which 
energy flux we denote $F(\phx)$, so that 
$F(\phx) \d \phx$ is the energy flux in the interval $\d \phx$. 
The photon energies, in units of 
the electron rest mass $\me c^2$, are denoted as 
$\phx$ and $\phe$ for photons of low ($<1$) and high ( $>1$) energies, respectively.
We define also  the power-law index $\alpha \equiv - \d \log F(\phx)/\d\log \phx$.

The radiation  field in AGNs has two major components.
The first component originates in the  accretion disc around 
the central source and 
its photons propagate along the jet direction. 
We consider a standard accretion disc \citep{ss73}, assuming 
a simple power-law $T(R)\propto R^{-3/4}$ dependence 
of temperature on radius, with the ratio of the outer 
to inner disc radius $\Rout/\Rin=10^4$ and the maximum 
temperature of $T = 5$ eV. The dimensionless temperature is then
$\Theta\equiv kT/ \me c^2 \approx 10^{-5}$. 
The resulting multicolor disc spectrum $\Fd(\phx)$ 
has a power-law shape with energy index $-1/3$ below $\Theta$ and 
has a Rayleigh-Jeans part at energies below 
$\Theta_{\min}= \Theta \ (\Rout/\Rin)^{-3/4} \approx 10^{-8}$.

 The second, isotropic component consists of the disc radiation
  scattered and reprocessed  in the broad line
region (BLR) and the infrared radiation by the dust \citep{s94,s96}. 
We take its  energy density  to be a fraction $\etaiso$ of the 
 energy density of the direct disc radiation, and its flux to be a cutoff power-law 
$\Fiso(\phx)\propto  \phx^{-\alphaiso} \exp(-\phx/\xmax)$ 
extending from the far-infrared $\xmin\sim10^{-9}$ 
to the UV band, as we assume $\xmax= \Theta$. 

The jet kinetic luminosity scales with the disc luminosity as 
\be 
\Ljet=\etaj \Ldisc = \Gamma\ \dot{M}c^2.
\ee
From the mass conservation (for a two-sided jet) 
$\dot{M}=2\pi \theta^2 R^2 \mp c \np(R)$, 
we get the Thomson optical depth across the jet  from the electrons associated 
with protons 
\be \label{eq:taut}
\taut(R)= \np(R)\sigmat \Rj= 2.4\ 10^{-5} \etaj \Ld /(R_{17}\theta \Gamma),
\ee
where we use standard notations $Q=10^x Q_x$ in cgs units.

Comoving value of the magnetic field in the jet is $\Bj$, its direction is
transversal by assumption. 
A reasonable scaling for the jet field
is $\Bj \propto 1/\Rj$, then the Poynting flux carried by the jet
\be
\Lb = \etab\Ldisc= {\Bj^2 \over 8 \pi} 2 \pi \Rj^2 \Gamma^2 c \sim 8\times 10^{43}
\Bj^2 R_{17}^2 (\theta\Gamma)^2 \ \mbox{erg\ s}^{-1} 
\ee
is constant along the jet.
Parameter  $\etab$ defines the role of the magnetic field 
and we consider two cases with low and high $\etab$.

For our analysis we consider two reference frames: 
the ambient medium (external) frame  and 
the frame comoving with the fluid with Lorentz factor $\Gamma$ (jet frame). 
Energies are normally given in the external frame.
The jet comoving values are specified using subscript 'c' or a prime.

\section{Step-by-step description of the breeding cycle}
\label{sec:anal}

\begin{figure}
\centerline{\epsfig{file=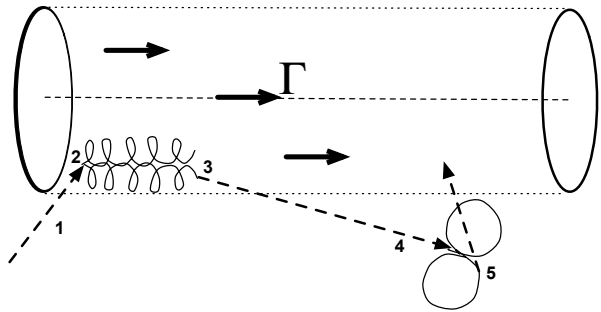,width=8.2cm}}
\caption{The scheme of the photon breeding cycle. The cylindrical jet  moves to
the right with Lorentz factor $\Gamma$. The size of the Larmor orbit is highly exaggerated.
See Section \ref{sec:anal} for details.}
\label{fig:scheme}
\end{figure}

The solution of the problem of the jet interaction with the external environment and 
dissipation of its kinetic energy into high-energy radiation requires full-scale nonlinear 
Monte-Carlo simulations as described in Section \ref{sec:numer}.
However, it would be useful to perform a simplified analytical study.
Very schematically, the process can be split into five steps (see Fig.~\ref{fig:scheme}).

%%%%%%%%%%%%%%%%%%%%%%%%%%%%%%%%%%%%%%%%%%%%%
%%%%%%%%%%%%%%%%%%%%%%%%%%%%%%%%%%%%%%%%%%%%%

 %\begin{enumerate}[\arabic]
% \item 
(1) A high-energy external photon (which origin is not important)
enters the jet and interacts with a soft photon producing an electron-positron pair.

(2) In the jet frame, the produced $e^+$ and $e^-$  originally move backwards 
relative to the direction of jet propagation with the energy $\sim \Gamma$
times higher than the energy of the parent photon (in the external frame).
When particles gyrate in the  magnetic field of the jet,
their time-averaged energy,  as measured in the external frame, 
becomes $\sim \Gamma^2$ higher than the parent photon energy.

%\item 
(3) The pair Comptonizes soft photons up to high energies.

%\item 
(4) Some of these photons leave the jet and produce pairs in the external
environment.

%\item 
(5) Pairs gyrate in the external environment and Comptonize soft photons more or less
isotropically. Some of the Comptonized high-energy photons enter the jet again.
This step completes the cycle.
%\end{enumerate}

Steps 1, 3, 4 and 5 require a field of soft photons to provide the conversion of 
the high-energy photons into pairs and to produce   new high-energy photons 
through the inverse Compton scattering. In our case, 
the direct disc radiation provides the opacity for the high-energy photons at
step 1 and provides a target for Compton scattering at step 5. 
Because the disc radiation is directed along the jet, 
its effect at steps 3 and 4 is small (interactions are suppressed by
the   factor $\sim 1/\Gamma^2$). 
The isotropic soft photon component provides the major source 
of seed soft photons for Comptonization at step 3, provides opacity at step 4
as well as contributes to the pair production at step 1.
 
An additional condition for high efficiency of the cycle 
is the presence of a transversal or chaotic magnetic field,
both in the jet (to provide step 2) and in the external environment (to provide
isotropization of pairs and photons Comptonized by them at step~5). 
 
Each step can be characterized by its energy transmission coefficient
$c_i$ defined (in the external frame) as the average ratio of total energy of particles with energy
above the pair production threshold  before and after the step.
 $c_2$ is large ($\sim \Gamma^2$), others are smaller than 1.
If the criticality index $\overline{C} \equiv c_1 \times c_2 \times c_3 \times c_4 \times c_5 > 1$,
 then the regime is supercritical, i.e. each cycle produces more particles than the previous one
and their number grows exponentially. In this case we deal with  particle
breeding rather with particle acceleration.  The spectrum of particles
changes slowly (and in principle the mean particle energy can decrease), 
but the number and the total energy of particles grow rapidly.

Each coefficient $c_k$ can be represented as an average of energy dependent
coefficient $C_k(\phe)$ over the photon spectrum at $k$-th step: 
\begin{equation}
c_k = \int {C_k(\phe) F_k(\phe) \d\phe,}
\end{equation}
where $F_k(\phe)$ is the energy flux spectrum at $k$-th step normalized to unity.
The spectrum  is a matter of a self-consistent treatment of the whole  
breeding cycle and we cannot estimate it {\it a priori}.
For simplicity, we estimate $C_k(\phe)$ for the constant external photon field 
(disc as well as isotropic).

A quantitative example for $C_k(\phe)$ behaviour
described in this section is given for the same parameters as in the detailed 
simulation presented in Section \ref{sec:example1}:  
$\Ld=1$, 
$R_{17} = 2$,  $\Gamma=10$, $\theta=0.05$, 
$\etaj=1$, $\etab=0.01$, 
$\etaiso=0.05$, and we consider here two cases of the
isotropic photon spectrum with $\alphaiso = 0.4$ and 1. 
The value $\alphaiso = 0.4$ seems reasonable for the BLR region, where 
the scattered radiation should be dominated by the UV and optical component and 
some less energetic IR radiation of dust is expected, while 
$\alphaiso = 1$  could be reasonable at the parsec scale where  dust radiation can be very 
important.

\subsection{Steps 1 and 4: propagation of photons and pair production}
\label{sec:step14}

Let us consider a problem of photon propagation through the boundary 
between the external medium and the jet.  
Let $\absgg(\phe)$ be the total opacity for pair production 
(i.e. the absorption coefficient produced by the disc and isotropic components) 
for a photon of energy $\phe$. 
The opacity produced by the isotropic component of the radiation field only
is denoted $\absiso  (\phe)$.

The jet photons move within the cone of opening angle $1/\Gamma$ to the 
jet direction and 
interact mostly with the isotropic component of the external radiation field,
 because the probability of interaction with the disc component  
is reduced by the  factor $\sim1/\Gamma^2$. The corresponding opacity 
computed for the interval perpendicular to the jet direction
is $ \aj \approx \Gamma\absiso (\phel)$. 
The probability that the jet photon  is  absorbed 
in the interval $\d\varpi$ at distance $\varpi$ from the jet boundary is 
exponential $\exp(-\aj \varpi) \aj \d\varpi$ (this assumes that all photons 
cross the boundary at the same angle $1/\Gamma$). 
The pair production and consequent Compton scattering transform 
this photon to photons of energy $\phei$.
Roughly half of these photons are emitted towards the jet and
the fraction $\exp(-\ai \varpi)$ reaches the boundary 
(where  $\ai= \absgg(\phei)$ is the absorption opacity and 
we assumed that the photons propagate in the direction 
perpendicular to the boundary). The probability 
that the photons are absorbed within the jet is about $1 - \exp(-\ai \Rj)$.
Assuming that only photons born at  a distance not 
more than the jet radius from the boundary can enter the jet,
we get the total  probability for an external photon  of energy $\phei$ 
to interact inside the jet with a soft photon 
producing an electron-positron pair: 
\beq \label{eq:c1}
C_1(\phei,\phel) &\approx  & 
\int_0^{\Rj} \rme^{ -\ai \varpi} \frac{1}{2}\rme^{-\aj \varpi} \aj\d \varpi  
\left[ 1 - \rme^{ - \ai \Rj} \right] \nonumber \\
&= &  \frac{1}{2} \frac{b}{1+b}  \left[ 1 -  \rme^{ -\ai \Rj} \right] \left[ 1 - \rme^{ - (\ai+\aj) \Rj} \right], 
\eeq
where 
\be 
b = \frac{\aj}{ \ai}= 
\frac{\Gamma \absiso  (\phel)}{ \absgg (\phei) } .
\ee
Obviously, $C_1$ depends also on the energy of the parent photon $\phel$, because 
it defines the spacial distribution of sources of photons $\phei$.

Analogously we derive the probability of a high-energy photon $\phel$ produced 
in the jet to escape from the jet and to produce a pair in the external medium at a distance not 
more than the jet radius from the boundary:
\beq \label{eq:c4}
C_4(\phel,\phei)  &\approx  & 
\int_0^{\Rj} \rme^{ -\aj \varpi} \frac{1}{2}\rme^{ -\ai \varpi} \ai\d \varpi  
\left[ 1 - \rme^{ -\aj\Rj} \right] \nonumber \\
&= &  \frac{1}{2} \frac{1}{1+b}  \left[ 1 - \rme^{ -\aj\Rj} \right] \left[ 1 - \rme^{ -(\ai+\aj) \Rj} \right] .
\eeq
Again, $C_4$  depends  on the energy of the parent photon $\phei$ that gave rise to the photon 
$\phel$ in the jet. 

If opacity is sufficiently high (i.e. we can ignore exponential factors in equations 
(\ref{eq:c1}) and (\ref{eq:c4})),  the sum of the probabilities $C_1(\phei,\phel) +C_4(\phel,\phei) =1/2$,
and their product  reaches the maximum of $1/16$ when the opacities are equal, $b=1$. 
Accounting for the angular distribution of photons gives
\be \label{eq:pmax}
\max \left\{ C_1(\phei,\phel)  \ C_4(\phel,\phei)  \right\} \approx 1/21.
\ee 

\begin{figure}
\centerline{\epsfig{file=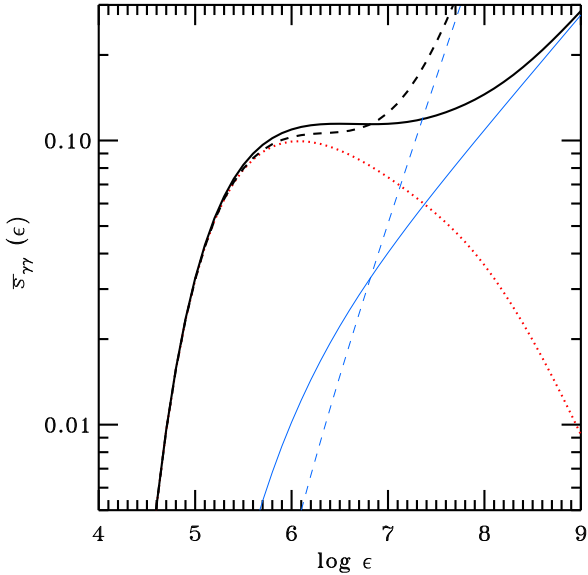,width=7.8cm}}
\caption{
The averaged over the photon distribution photon-photon absorption 
cross-section $\overline{s}_{\gamma\gamma}(\phe)$ 
given by equation (\ref{eq:oversgg}). 
The dotted  curve gives the cross-section averaged over the multicolor disc spectrum
with $\Theta =10^{-5}$ and $\Rout/\Rin=10^4$. 
Almost straight lines give the opacity induced by the isotropic component 
of the radiation field  $\propto \phe^{\alphaiso} \exp(-1/\phe\Theta)$
(scaled by the ratio
of photon densities $N_{\rm ph, iso}/N_{\rm ph, d}$ 
of the isotropic and disc radiation).
Thick curves give the sum of the two opacities. 
Solid and dashed lines are for 
$\alphaiso=0.4$ and   $\alphaiso=1$, respectively.
}
\label{fig:alphagg}
\end{figure}

Let us now discuss in details the behaviour of the photon opacity. 
For isotropic external photons, the main contribution to the photon opacity 
 is given   by the radiation of the accretion disc $\Fd(\phx)$.
The  angle-averaged  cross-section of photon-photon pair production 
$\sigma_{\gamma\gamma}(\phe, \phx)$ has  
a low-energy threshold $ \phx \phe >1$ and 
reaches the maximum $\approx 0.21\sigmat$ at  $\phx  \phe \approx 3.5$
\citep[see e.g.][]{zdz88}.  
The absorption coefficient 
for a photon of energy $\phe$ can be obtained by 
integrating the cross-section over the spectrum of soft photons: 
\be \label{eq:tau}
\absgg (\phe)= 
N_{\rm ph} \sigmat \overline{s}_{\gamma\gamma}(\phe) ,
\ee
where the mean cross-section (in units of $\sigmat$)
\be \label{eq:oversgg}
\overline{s}_{\gamma\gamma}(\phe)= \int_{1/\phe}^{\infty} 
n(\phx) \sigma_{\gamma\gamma}(x\phe) \d \phx 
\ee
and $n(\phx)$ is the photon number density  normalized to unity.
The mean cross-section for the multicolor disc 
is plotted in Fig. \ref{fig:alphagg} by the dotted curve.
The total disc photon number density can  be expressed through 
the energy density $N_{\rm ph, d}=\xi U_{\rm d} / kT$, where   
$\xi \approx 1.4$ for the multicolor disc. Because 
$U_{\rm d} = 2 \Ldisc /(c 4 \pi R^2)$ close to the jet axis
(factor 2 comes from the angular distribution of the disc radiation, which 
we assume to follow the Lambert law), we obtain
\beq \label{eq:alphagg}
\absgg (\phe) & = &  
\frac{\Ldisc}{c4\pi R^2} \frac{\sigmat}{\me c^2}  
\frac{1}{\Theta} \ 2 \ \xi \ \overline{s}_{\gamma\gamma}(\phe)\nonumber  \\
& = &  6 \ 10^{-14}  \frac{\Ld}{ R^2_{17} \Theta_{-5} } 
\overline{s}_{\gamma\gamma}(\phe)  \ \mbox{cm}^{-1}.
\eeq
The opacity is low at $\phe < 1/\Theta \sim 10^5$ and has 
a maximum at $\phem \approx 10/\Theta$: $\absggmax =  
 6 \ 10^{-15} \Ld/( R^2_{17} \Theta_{-5})  \ \mbox{cm}^{-1}$.
At higher energies $\phe >\phem$, the photons interact with the power-law part 
of the multicolor disc spectrum resulting
%which can be represented as 
%\be 
%\Fd(\phe) \approx \frac{20}{\pi^5} \frac{\Ldisc}{R^2} 
%\frac{1}{T} \left( \frac{\phe}{T} \right)^{1/3} ,
%\ee
in a power-law decay of the opacity: 
\be 
\absgg (\phe)\approx   \absggmax
\left( \frac{\phem}{\phe} \right)^{1/3},
\ee
which transforms into a faster $\frac{1}{\phe}\ln \phe$ decay at $\phe>1/\Theta_{\min}$.

%%%%%%%%%%%%%%%%%%%%%%%%%%%%%%%%%%%%%%%%%%%%%
%%%%%%%%%%%%%%%%%%%%%%%%%%%%%%%%%%%%%%%%%%%%%

In spite of a lower energy density, 
the isotropic component $\Fiso(\phx)$ dominates the opacity at very high $\phe$,
because of its much softer spectrum $\alphaiso>-1/3$.
The opacity increases with energy following a power-law  \citep{s87}
\be 
\absiso (\phe) \approx \frac{1}{5}   
\frac{\sigmat}{\me c^3}   \Fiso(1/\phe) \propto \phe^{\alphaiso},
\ee
as shown in Fig. \ref{fig:alphagg}.

Because the coefficients $C_1$ and $C_4$ depend on both 
photon energies $\phei$ and $\phel$, it is easier to visualize them as
one-dimensional cross-sections. 
As an example, we plot $C_1(\phei,10\phei)$ and $C_4(\phel,\phel/10)$ in Fig. \ref{fig:coefs}. 
Coefficient $C_1$ vanishes  below $\phei\ll \phem$, because of low opacity when the 
photons escape freely and do not produces pairs within the jet.
In the interval $5\ 10^4<\phei<10^6$,  $C_1\sim b/2(1+b)\sim 1/4$ because here 
the ratio of opacities $b\sim 1$.  
At higher energies, the opacity is dominated  by the isotropic component,  
$b\approx \Gamma \absiso(\phel)/\absiso(\phei) \sim \Gamma (\phel/\phei)^{\alphaiso}$
is large and $C_1\sim 1/2$.

Coefficient $C_4$ vanishes at $\phel<2\ 10^5$ because the opacity 
produced by the isotropic  power-law component  is low here, and photons 
are escaping too far from the jet (note the exponential factor in eq. [\ref{eq:c4}]).
At $\phel\sim10^6$, $b\sim1$ and $C_4\sim1/4$. At high energies $C_4$ decays 
$\propto 1/2b$.
In our formulation of $C_4$, we accounted only for photons that can directly  
penetrate from the jet to the external medium. However, those high-energy photons that are
absorbed within the jet do not leave the system, but  produce pair-photon cascade. 
During the cascade the photon energy 
eventually becomes sufficiently low to allow the photon escape from the jet 
and to produce a pair outside. Thus even if a photon was emitted towards the jet axis, 
a fraction of its energy can eventually escape the jet in a form of secondary photons.

\subsection{Step 2: energy gain of a pair produced in the jet}
\label{sec:step2}

A high-energy photon of energy $\phei$ at this stage has interacted 
with a  soft photon inside the jet to produce an electron-positron pair.
The two particles move now backwards relative to the 
jet propagation  direction with the mean comoving Lorentz factor 
\be \label{eq:gammac}
\gammac=\frac{4}{3} \frac{\phei}{2} \Gamma.
\ee
Here the factor 4/3 comes from the averaging 
over the angles of the incident high-energy photons accounting for the probability 
of interaction with the disc photons moving along the jet (in case of interaction with 
the isotropic photon field, this factor disappears). 
Integrating over the Larmor orbit, we get 
the mean particle energy in the external frame 
\be 
\langle \gamma \rangle=  \gammac \Gamma\  \langle (1+ \cos\phic)^2 \rangle_{\phic} 
  = \frac{3}{2} \gammac \Gamma = 2 \frac{\phei}{2} \Gamma^2,  
\ee
where $\phic$ is the angle between the particle momentum and jet propagation 
direction (in jet frame).  Note, that the averaging is over $\phic$, not 
$\cos \phic$.
The mean energy gain is then 
\be
C_2 (\phei) = \frac{\langle \gamma \rangle}{\phei/2}  = 2 \Gamma^2. 
\ee
 
\begin{figure}
\centerline{\epsfig{file=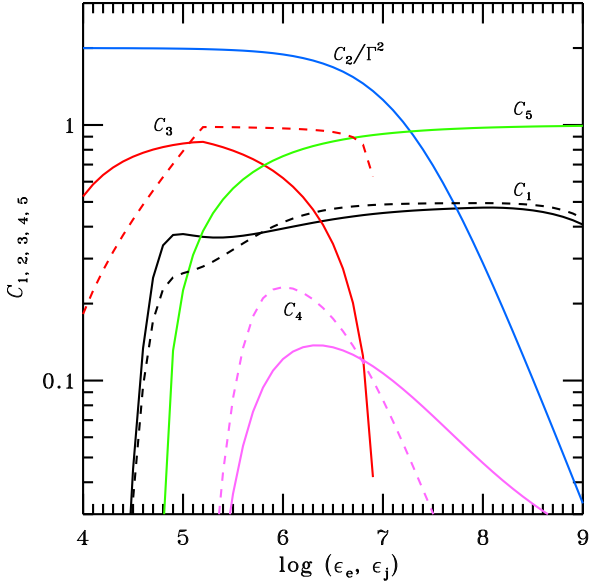,width=7.8cm}}
%\begin{center}
%\leavevmode
%\epsfxsize=7.8cm \epsfbox{fig3a_coefficients.ps}
%\hspace{0.5cm}
%\epsfxsize=7.8cm \epsfbox{fig3b_coefficients.ps}
%\end{center}
\caption{Coefficients $C_i(\phe)$ versus photon energy. 
$C_2$ and $C_3$ are functions of energy of a photon produced 
in the external medium $\phei$, while $C_5$ is the function of
a typically higher photon energy $\phel$,  produced in the jet.  
Coefficients $C_1$ and $C_4$ depend on both energies, and 
we plot here only the one-dimensional cross-sections  
$C_1(\phei,10\phei)$ and $C_4(\phel,\phel/10)$.
Solid curves represents the case $\alphaiso=0.4$, while dashed curves are for  
$\alphaiso=1$ ($C_2$ and $C_5$ do not depend on the choice of 
$\alphaiso$).
}
\label{fig:coefs}
\end{figure}

The time-scale of synchrotron cooling is equal to 
the inverse of Larmor frequency at the (comoving) electron Lorentz 
factor 
\be \label{eq:gmax}
\gmaxc \approx 10^8 { \Bj^{-1/2}} .
\ee 
Compton losses because of the disc radiation can be neglected at this step 
(for our parameters) 
due to a deep Klein-Nishina regime of scattering at $\gammac \gtrsim 10^{8}$.
For very high initial photon energy $\phei \gg \gmaxc  \Gamma$, 
the Lorentz factor of  the produced pair  is  reduced to $\sim \gmaxc$ 
after first Larmor orbit because of the synchrotron losses.
 In the external frame its energy is now $\sim \gmaxc \Gamma$,
resulting in the energy gain (loss) factor $C_2(\phei)\approx \gmaxc\Gamma/\phei$. 
Thus the energy gain for arbitrary $\phei$  (shown in Fig. \ref{fig:coefs}) 
can be written as
\be \label{eq:c2}
C_2(\phei)=  2 \Gamma^2 \frac{1}{1+\phei \Gamma/\gmaxc} .
\ee

%%%%%%%%%%%%%%%%%%%%%%%%%%%%%%%%%%%%%%%%%%%%%
%%%%%%%%%%%%%%%%%%%%%%%%%%%%%%%%%%%%%%%%%%%%%

\subsection{Step 3: energy conversion into photons}
\label{sec:step3}

Coefficient $C_3(\phei)$ describes a fraction of the electron (positron) energy,
which is converted into photons with the energy above the thresholds of the cycle  
$\phelmin= 3\ 10^5$ defined by a sharp drop of coefficient $C_4$. 
Photons of lower energies  escape almost freely from the jet and are lost 
from the cycle.
$C_4$ also drops above  $\sim3\ 10^7$, however, 
the photons of higher energies are not lost from the cycle, but produce a pair cascade and 
eventually escape from the jet producing pairs outside.

The efficiency of conversion is mainly determined  by the ratio between Compton and synchrotron losses
(because synchrotron photon energy is below $\phelmin$ in our case), which,
in Thomson regime,  is just the ratio of the soft radiation energy density 
(dominated by the isotropic component) to the magnetic field energy density
(measured in the jet frame): 
\be \label{eq:urub}
\frac{\Urad'}{\UB '} = {\etaiso U_{\rm d} \Gamma^2 \over \Lb /(2\pi \Rj^2 \Gamma^2 c)} =
{\etaiso\over \etab} (\Gamma\theta)^2 \Gamma^2 .
\ee
In reality, however, the share of Compton 
scattering is much smaller, because a pair produced in the jet by 
a photon at  $\phei \sim 1/\Theta $ gains energy up to $\gamma \sim \Gamma^2/\Theta$ and 
interacts with photons at $\phx \sim \Theta $ in the deep Klein-Nishina regime. 
The situation also complicates by the fact, that because of cooling, the 
fraction of soft photons interacting with the pair in Thomson regime 
grows with time.

Neglecting for simplicity scattering in the Klein-Nishina regime and 
pair cooling (i.e. assuming constant pair Lorentz factor given by Eq. [\ref{eq:gammac}]), 
we define  coefficient $C_3$ as the ratio 
of the soft radiation energy density, that can produce photons above 
the threshold, to the sum of the magnetic field energy density and 
the total radiation density in the Thomson regime: 
\be \label{eq:c3def} 
C_3(\phei) =  \frac {\Uc' }{ \UB'  + \UT '} ,
\ee
where
\beq \label{eq:ub}
\UB'  & =& \Bj^2/(8\pi), \\
\label{eq:ut}
\UT' &  \propto & \Gamma^2 \int_{\xmin}^{ \xkn} \Fiso (\phx) \d \phx , \\
\label{eq:uc}
\Uc'  & \propto &  \Gamma^2 \int_{\max[\xmin,\xlow]}^{\xkn} \Fiso (\phx) \d \phx   . 
\eeq
The upper integration limits is defined by the Thomson regime of scattering
\be \label{eq:xkn}
\phx < \xkn = \frac{1}{\langle\gamma\rangle} = \frac{1}{\phei\Gamma^2} .
\ee
The lower limit is given by the condition that 
the scattered photon energy $\phel=   \phx  \langle\gamma^2 \rangle$ 
is above the threshold $\phelmin $.
Averaging over the Larmor orbit, we get the mean square of the particle Lorentz factor:
\be 
\langle \gamma^2 \rangle=  \gammac^2 \Gamma^2 \  \langle (1+ \cos\phic)^3 \rangle_{\phic} 
  = \frac{5}{2} \gammac^2 \Gamma^2 =  \frac{10}{9} \phei^2 \Gamma^4 ,
\ee
that gives 
\be \label{eq:low}
\phx  > \xlow = \frac{\phelmin}{ \langle\gamma^2 \rangle }  = 
 \frac{9}{10} \frac{\phelmin} {\phei^2 \Gamma^4 } . 
\ee
The behaviour of $C_3(\phei)$ is  shown in Fig. \ref{fig:coefs}.
 
There are three characteristic energies. 
For $\phei < \phe_1 \approx \phelmin /\Gamma^2\sim  10^3$, 
 $ \xkn  <  \xlow$ and $C_3$ vanishes. 
 At energies  $\phei>\phe_2  \approx \sqrt{\phelmin / \xmin} /\Gamma^2 \sim 10^5$,
 $\UT =\Uc $ because $ \xlow < \xmin$. 
And at energies above $\phe_3= 1/(\Gamma^2 \xmin) \sim 10^7$,  $\xkn<\xmin$ 
and $C_3=0$ in our approximation. 
At  energies  $\phe_2<\phei<\phe_3$, one can approximate 
\be \label{eq:c3app} 
\frac{1}{C_3(\phei) }  \approx
\left\{ \begin{array}{ll} \strut\displaystyle
1 +  \frac{\UB' }{\Urad'}   \frac{ \xmax^{1-\alphaiso} -  \xmin^{1-\alphaiso}  } 
{ \xkn^{1-\alphaiso} -  \xmin^{1-\alphaiso}   }   , & \alphaiso\ne1, \\
\strut\displaystyle
 1+ \frac{\UB '}{\Urad'}    \frac{ \ln(\xmax/\xmin) }{ \ln(\xkn/\xmin) }, &  \alphaiso=1.
\end{array} \right. 
\ee

 Taking into account equation (\ref{eq:urub}),
we can conclude that  Comptonization at step 3 can be very efficient ($C_3\sim1$) if:
\begin{enumerate}
\item The magnetic field is weak, $\etab \ll 1$. 
\item  The energy density  of the external  isotropic radiation is large in the jet frame, 
i.e. $\Gamma$ or $\etaiso$ are large. 
\item The spectrum of isotropic component is soft, $\alphaiso \gtrsim 1$. Then the power-law 
decay of $C_3$ above $\phe_2$ is replaced by a constant
or a logarithmic decay   (see eq. [\ref{eq:c3app}] and notice a slow decays of $C_3$ in 
Fig. \ref{fig:coefs}, dashed curve).
In this case, a high value of $C_3$ is possible even 
for magnetically dominated jet if its total power 
is a few times smaller than the disc luminosity. 
\end{enumerate}

\subsection{Step 5: Compton scattering in the external environment}
\label{sec:step5}

At step (5) an electron-positron pair, produced in the external 
enviroment, Comptonizes
a number of soft photons to produce high-energy photons above the energy threshold  
$\pheimin \sim 3 \ 10^4$ defined by the low-energy cutoff of $C_1$.  
 At this step, synchrotron losses can be safely neglected as 
 magnetic field in the external medium is much lower than   the jet field.

%%%%%%%%%%%%%%%%%%%%%%%%%%%%%%%%%%%%%%%%%%%%%
\begin{figure}
\centerline{\epsfig{file=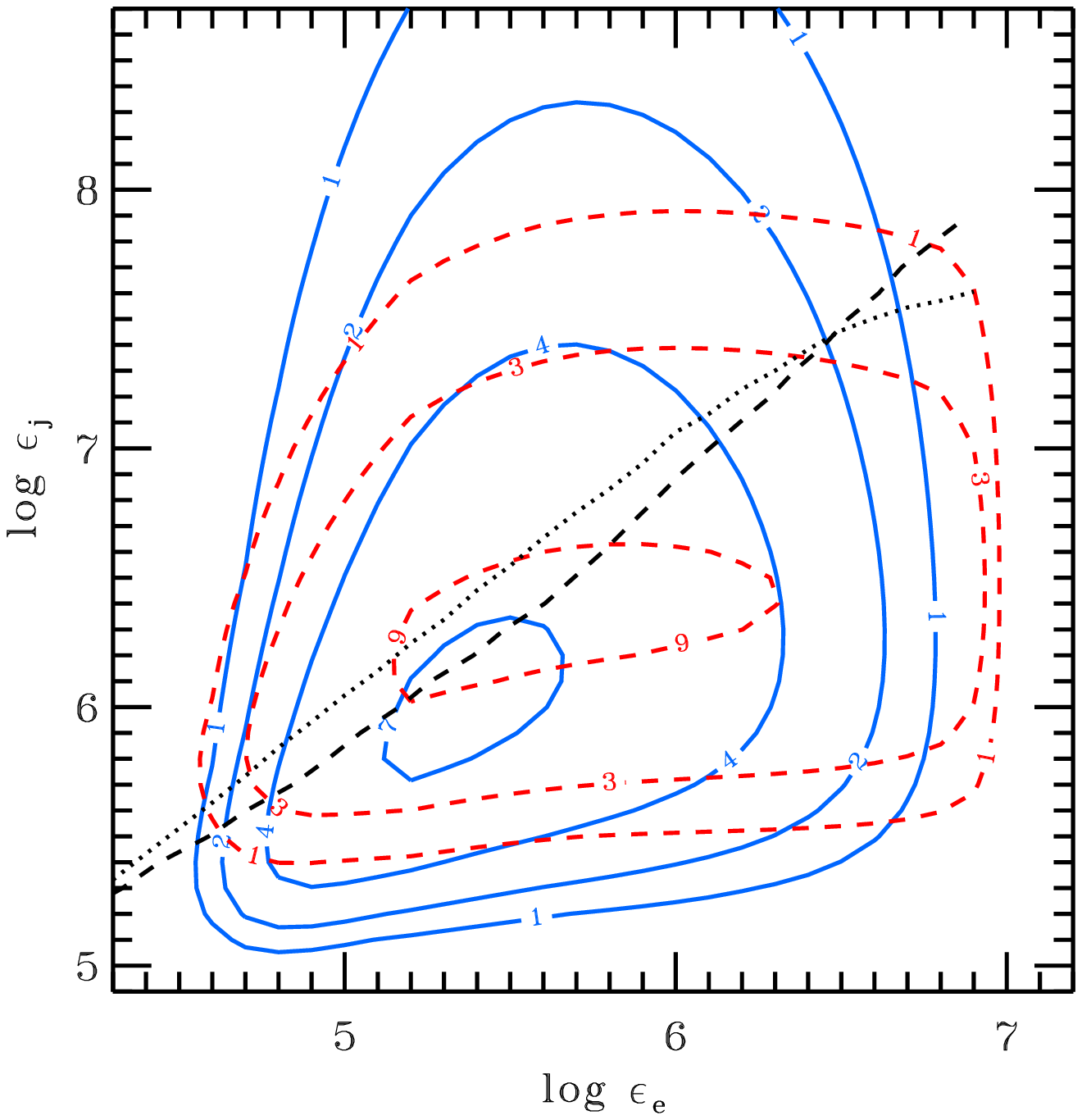,width=7.8cm}}
\caption{The product of the five coefficients $A(\phei,\phel)= 
C_1(\phei, \phel) C_2(\phei) C_3(\phei) C_4(\phel,\phei) C_5(\phel)$,
represented by contours of constant levels. 
For  $\alphaiso=0.4$ (solid contours), it reaches the maximum  of 7.4 at  
$\phei = 2\  10^5$, $\phel = 10^6$. 
In this case, the dotted curve shows the median energy $\phel(\phei)$ for steps 2 and 3, and
the dashed curve shows the median energy  $\phei(\phel)$ at step 5. 
For $\alphaiso=1$ (dashed contours),  the maximum of 9.9
is at $\phei = 3\ 10^5$, $\phel = 2\ 10^6$. 
}
\label{fig:map}
\end{figure}
%%%%%%%%%%%%%%%%%%%%%%%%%%%%%%%%%%%%%%%%%%%%%

Because cooling is fast, the spectrum produced   by a pair is 
a standard ``cooling'' spectrum $F(\phe) \propto \phe^{-1/2}$ extending
up to the maximum energy  equal to the pair Lorentz factor $\gamma=\phel/2$. 
Defining coefficient $C_5(\phel)$ as the energy fraction emitted 
above the thresholds $\pheimin$, we get
 \be \label{eq:c5}
 C_5(\phel) = 1- \left( \frac{\pheimin}{\gamma} \right)^{1/2}   = 
 1- \left( \frac{2 \pheimin}{\phel}   \right)^{1/2}  . 
 \ee
The computed $C_5(\phel)$ is shown in  Fig.  \ref{fig:coefs}. 
A  low-energy cutoff is at $\phel=2  \pheimin  =6\ 10^4$.

\subsection{Amplification through the cycle and dissipation efficiency}

The absolute theoretical maximum of the amplification factor 
\be 
\max {A} (\phei,\phel) = C_1 C_2  C_3 C_4  C_5 
\approx \Gamma^2 / 10 
 \ee  
 is achieved when  $C_2=2 \Gamma^2$, $C_1 \times C_4 = 1/21$, and
 $C_3=C_5=1$.
Thus, the minimum jet Lorentz factor required to achieve 
supercriticality is $\Gamma \approx 3$.
In a more realistic situation, the amplification factor is smaller and larger $\Gamma$ is 
needed.  For example, strong magnetic field inhibits the photon 
breeding reducing $C_2$ and $C_3$ because of synchrotron losses.
The cascade still develops, if a sufficiently dense soft external radiation 
field is present, because $C_3\sim1$, when $\etaiso\Gamma^4/\etab$ is large
(Section \ref{sec:step3}), see Sect. \ref{sec:example2} for the simulation example.

An estimations of $A$ using our simple formulae from Sections 
\ref{sec:step14}--\ref{sec:step5}  is shown in Fig. \ref{fig:map}. 
For $\alphaiso=0.4$, it reaches 7.4 at  $\phei = 2 \ 10^5$, $\phel =  10^6$ 
(the more precise Monte-Carlo step-by-step simulations of the cycle
efficiency show $A$ reaching the maximum of 4.7  at  $\phei =  5\ 10^5, \phel = 2\ 10^6$). 
For  $\alphaiso=1$ the maximum  of $A$ is about 9.9 at  $\phei = 3 \ 10^5$, $\phel = 2 \ 10^6$.
 Of course,  $A(\phei,\phel)$  does not have the meaning of the criticality index $\overline{C}$.
The photon energy  distribution  during the cycle is wide and does not necessarily 
coincide with the area of the maximal amplification. 
Fig. \ref{fig:map}  (dotted curve) shows the
 median energy $\phel(\phei)$ computed by a Monte-Carlo method 
  (for  $\alphaiso=0.4$)  of the distribution   $\phel p_{23}(\phel,\phei)$, where 
$p_{23}(\phel,\phei)$ is the probability density of a photon 
of energy $\phei$ to produce a photon of energy $\phel$
after steps 2 and 3. A similar function $\phei(\phel)$  for step 5,
which is a median of distribution $\phei p_5(\phei,\phel)$, is also shown (dashed curve).
Note that the median energy gain at 
steps 2 and 3 is slightly higher than the energy loss at step 5, which means 
that photon energy rises on average during the cycle  and tends to the area
at  $\phei \sim 3\ 10^6$, $\phel \sim 3\ 10^7$,
where the curves intersect. The amplification in that area 
is $A(\phei,\phel) \approx 2$. 

We can now estimate a typical distance where the most efficient 
energy dissipation in the shear flow should take place. 
The total optical depth across the jet is (see eq. [\ref{eq:alphagg}])
\be 
\taugg (\phe) = \absgg(\phe) \Rj
= 300 \frac{\Ld (20\theta)}{ R_{17} \Theta_{-5} } 
\overline{s}_{\gamma\gamma}(\phe)  ,
\ee
which gives us the maximal  distance from the central engine where 
$\taugg (\phem) \sim 1$ and, therefore, is sufficient to produce pairs in the jet:
\be \label{eq:rmax}
R_{\max} \sim  \frac{\Ld  \ (20\theta) }{  \Theta_{-5} }\ \mbox{pc}.
\ee 
Note that infrared emission of  the dust can provide the opacity for higher energy photons 
at much larger distances. 

At $R<R_{\max}$â the typical depth of photon penetration $\delta\varpi =\delta r\ \Rj$ 
into the jet  is given by the condition $ \delta\varpi\ \absgg (\phe)=1$, 
i.e. $\delta r= 1/\taugg (\phe)$. 
At the peak of absorption, for  $\phe\sim \phem$, we obtain 
\be  \label{eq:deltar}
\delta r = \frac{R}{R_{\max}} = 0.03 \frac{\Theta_{-5}R_{17}}{\Ld  \ (20\theta) } .  
\ee
The value of $2\delta r$  defines the fraction of the jet volume  occupied by the 
``active layer'', which is responsible for the 
exponential breeding of the high-energy photons. 
Because the photons of slightly higher and lower energies than $\phem$ 
can propagate further into the jet, the fraction of the total jet kinetic energy released is even higher 
that $2 \delta r$,  as demonstrated by  the full scale simulation (see Section \ref{sec:simul}).
At small distances from the central source, opacity is 
so large that high-energy photons can hardly penetrate the boundary between the jet 
and the external medium. The cascade can develop there, but the dissipation 
efficiency is very low because  $\delta r$ is small.

The efficiency also depends on the number of generations $N$ the breeding cycle operates. 
A high efficiency is achieved if $N$  is larger than a few (say  $N=10$). 
The time-scale of the cycle is defined by the longest step 4, which takes on average
$t_{\rm cycle} \sim \delta r\ \Gamma$. 
Because the active layer width depends on distance $\delta r \propto R$, 
the breeding cycle takes longer time (in $\Rj/c$ units) at larger $R$ and 
at the same time it is less sensitive to the requirement of a sharp boundary.
The cycle time-scale should be compared to
the dynamical time-scale $t_{\rm dyn} \sim 1/\theta$. 
This determines the most favourable distance for the breeding to operate as
\be 
R_{\rm eff} = \frac{R_{\max}}{N\Gamma\theta} =  
 \frac{1}{N}  \frac{\Ld}{ \Theta_{-5} }   \frac{20}{\Gamma} \ \mbox{pc}. 
 \ee

For this estimation we assumed that the amplification factor $A$ does not depend 
on distance $R$. This is true as long as $\Bj^2$,  $\Fd$ and $\Fiso$ all scale as $1/R^2$ 
and  $\delta r\ll 1$.
The last condition breaks down at a parsec scale (see eq. [\ref{eq:rmax}]), 
while the scaling for $\Fiso$  probably breaks 
much closer, as one can hardly expect the existence of the sufficient
material to provide efficient scattering/reprocessing of the disc radiation to the isotropic
component beyond the BLR. 
Thus, the most efficient dissipation is expected within the BLR 
at $R_{\rm eff} < R < R_{\max}$.

There still remains such source of $\Fiso$ as the dust IR radiation, but
it cannot convert photons at $\phel < 10^7$ into pairs and the 
coefficient $C_4$ will be small. In such a case one can expect that the cycle  
works in  the TeV range, which could be relevant for TeV blazars.
At small distances $R < 10^{16}$ cm, there appears another possibility:
nonthermal X-ray component of the disc radiation (which typically constitutes $\sim 0.1$ of the disc luminosity) becomes opaque and can efficiently convert lower energy photons 
($\phe < 1/\Theta$). Then one can expect the existence of a low-energy
breeding cycle in the range $\phe \sim 10^3$--$10^4$. This possibility requires a separate study.

\section{Numerical implementation}
\label{sec:numer}

We assume a sharp boundary between the jet and
the surrounding medium at the start of simulations. 
For simplicity we adopt the constant physical conditions
along the $20\Rj$ interval, including a uniform field of the external radiation.
The jet decelerates transferring its momentum to the radiation. 
In our model we take into account the dependence of the fluid 
Lorentz factor on $r$ only. 

%There should be dependence on $z$,
%of course, but its treatment would require the account of hydrodynamical effects. 
%Therefore we 
%are not able to reproduce a steady-state behaviour of the jet, which does 
%imply a $z$-dependence. For this reason we formulate our problem as a study of
%the system evolution starting from the non-radiating state with the constant 
%Lorentz factor through the jet volume.

The  numerical simulation method is based on the Large Particle
Monte-Carlo code (LPMC) developed by \citet{s85} and \citet{s95}. The code can treat
essentially nonlinear problems when the simulated particles constitute
at the same time a target medium for other particles. The number of large
particles (LPs) representing  photons and $e^\pm$ pairs was $2^{17} = 131072$.

%%%%%%%%%%%%%%%%%%%%%%%%%%%%%%%%%%%%%%%%%%%%%
\begin{figure}
\centerline{\epsfig{file=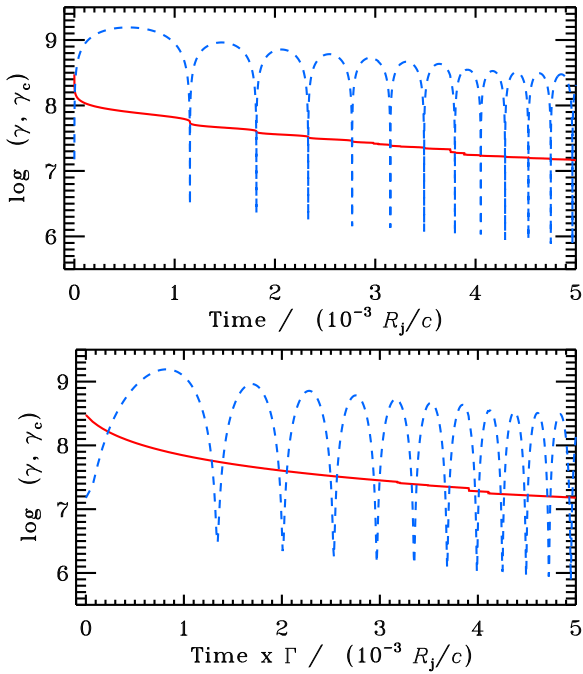,width=7.8cm}}
\caption{External frame ($\gamma$, dashed curve) and jet comoving ($\gammac$, 
solid line) Lorentz factor
of a particle gyrating in transversal magnetic field $\Bj = 0.35$~G frozen
into fluid. Upper panel: dependence  on the external frame time.
Lower panel: dependence on the comoving time multiplied by the bulk Lorentz factor $\Gamma = 10$.
Initial Lorentz factor of the particle is $\gammac = 3\  10^8$ and 
the jet radius $\Rj = 10^{16}$ cm.
}
\label{fig:orbit}
\end{figure}
%%%%%%%%%%%%%%%%%%%%%%%%%%%%%%%%%%%%%%%%%%%%%

The version of LPMC
used here treats Compton scattering, synchrotron radiation,
photon-photon pair production, and pair annihilation. Synchrotron self-absorption
was neglected as it consumes too much computing power and is not very important
in this application. All these processes are
reproduced without any simplifications at the micro-physics level.
The general organization of the LPMC simulation is described in \citet{s95}. 
A new specific feature of numerical simulation in this work (as well as 
in \citealt*{s03}) is a  scheme of particle tracking in the relativistic fluid.

Parameters of photon LPs are defined in the external reference frame.
The energy and the direction of electron/positron LPs are defined in the 
jet frame, since the electron energy in the external frame oscillates by a factor 
$4\Gamma^2$ because of gyration in the magnetic field. 

The tracking scheme for high-energy charged particle LPs differs 
for the first Larmor orbit and the rest of the trajectory.
The reason is that at the first orbit a particle can lose 
the main fraction of its (jet frame) energy before it 
turns around and gains the energy in the external frame 
(see Sect. \ref{sec:step2}  and Eq. \ref{eq:gmax}). 

Rapid energy losses require a fine particle tracking at the first Larmor orbit
if the particle has been produced with $\gammac > \gmaxc$.
Therefore, the comoving tracking step is limited by $\d s = 0.1 \RL$, 
where $\RL = 1.7 \ 10^3 \gammac/\Bj\ \mbox{cm}$ is the Larmor radius. 
Each step is described in both reference frames. The corresponding Lorentz
transformations from the jet frame to the external frame are:
\begin{equation}
\d t = \d t_{\rm c} \left( 1 + {V\over c}\betac \cos \phic\right) \Gamma, \ \  \gamma = 
\gammac \left ( 1 + {V\over c}\betac \cos \phic \right) \Gamma ,
\label{eq:lorentz}
\end{equation}
where $V$ is the velocity of the fluid, $\d t_{\rm c}$ is comoving time interval
for the tracking step, $\betac$ is the particle comoving 
velocity in units of $c$,  $\phic$ is the angle between 
comoving direction of the particle momentum and the jet, $\d t$ and $\gamma$ are 
the external frame values for the step time-interval and the particle Lorentz factor. 
The comoving representation is used to track the particle gyration in the 
magnetic field frozen into the jet and to simulate the synchrotron 
radiation. The external frame representation 
is more convenient  to simulate interactions with photons and is
necessary to synchronize the particle tracking with the general evolution of 
the system.  

An example of high-energy ($\gammac > \gmaxc$) particle 
tracking for several Larmor orbits
as viewed from both reference frames is shown in Fig.~\ref{fig:orbit}. The magnetic field 
and the external soft photon field correspond to the case considered 
in Sect. \ref{sec:example1}. One can see the dramatic energy loss due to synchrotron
radiation at the first orbit and a much slower further evolution. 
A discrete step at $t \sim 4\ 10^{-3}$ is a result of Compton scattering.

When the particle energy is below $\gmaxc$, we can neglect 
the dependence
between the comoving direction and the energy. In this case we sample the direction 
of the particle assuming a uniform distribution of its gyration phase 
$\phic$ in the comoving system. The external frame probability density function
for $\phic$ is 
\begin{equation}
p(\phic) \propto \d t/\d t_{\rm c} = \left(1 + {V \over c} \betac \cos\phic \right)\Gamma.
\end{equation}

Trajectories and momenta of LPs are three-dimensional. The target LP density
is averaged over 45 two-dimensional cylindric cells: 5 layers along the
jet with 9 concentric shells in each.
The trajectories of electrons and positrons in the magnetic field were
simulated directly assuming transversal geometry of the field $\Bj$
in the jet and $\Be$ in the external matter.

The primary soft photon field (disc and external isotropic) is kept constant during the simulations. 
Any additional soft (synchrotron) photons  produced by the cascade 
participate in the simulations in the form of LPs.
 At the start of simulations, the shell of the length $\Delta z=20$
 between  $10 < z < 30$, and the radial extend $0.9 < r  < 1$ is filled
by seed isotropic high-energy photons, whose energy density is several orders 
of magnitude less than the energy density of the jet. 
%After that 
%there is no additional injection of external photons 
%(except constant seed soft photons) and
All particles participating in the simulations are descenders of these
seed photons.
In the course of the simulations,  the jet undergoes  differential deceleration. 
We split the jet into 500 cylindric shells, calculate the momentum transferred to
each shell and decelerate each shell independently from others.  
%This is a very rough
%simplification: the deceleration should depend on $z$ and this will lead to a
%complicated velocity pattern including formation of internal shocks.

\section{Results of simulations}
\label{sec:simul}

 We have made several tens of simulation runs with different parameters and
various model formulations. 
In some of them we observed the exponential energy grow, others gave no effect. 
As we have shown in Section \ref{sec:anal}, the  supercritical behavior 
appears, when the jet Lorentz factor and the density of the soft isotropic radiation 
is sufficiently high. The lowest Lorentz factor, where the cascade is 
possible in principle is around  $\Gamma = 4$.  
Here we present only two examples demonstrating the development of the
runaway cascade for different conditions.

\subsection{Example 1. Weak magnetic field and a ``minimal'' seed radiation}
\label{sec:example1}

In this example we  assume the disc luminosity $\Ld=1$,
the distance of the active region from the black hole $R_{17} = 2$,  and
jet parameters $\Gamma=10$, $\theta=0.05$. 
The total jet kinetic power is equal to the disc luminosity 
 $\etaj =1$ and the Poynting flux is one per cent of that $\etab=0.01$. 
This implies a matter-dominated jet. 
The resulting  magnetic field in the jet  (comoving frame) is $\Bj\approx 0.35$~G
and we take  the external magnetic field $\Be=10^{-3}$~G.
The ratio of isotropic and disc radiation 
energy densities is $\etaiso = 0.05$, and we assume $\alphaiso=0.4$.

%%%%%%%%%%%%%%%%%%%%%%%%%%%%%%%%%%%%%%%%%%%%%
\begin{figure}
\centerline{\epsfig{file=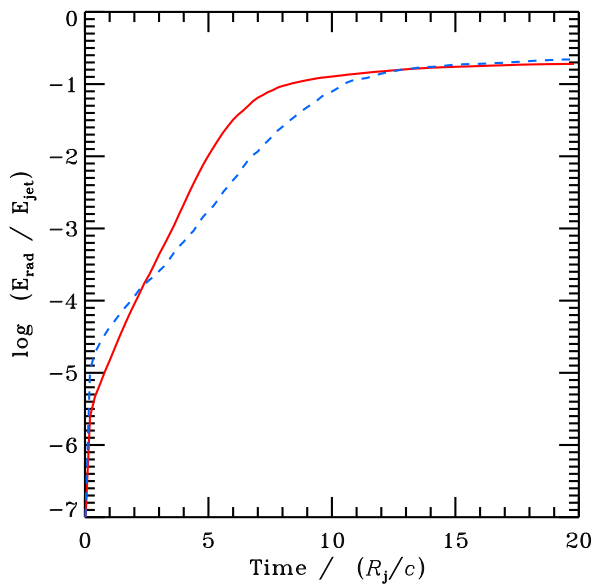,width=7.8cm}}
\caption{Total cumulative  energy release of the fluid
into radiation versus time.
Solid curve: Example 1; dashed curve: Example 2 (see text).}
\label{fig:evol}
\end{figure}
%%%%%%%%%%%%%%%%%%%%%%%%%%%%%%%%%%%%%%%%%%%%%

\begin{figure}
\centerline{\epsfig{file=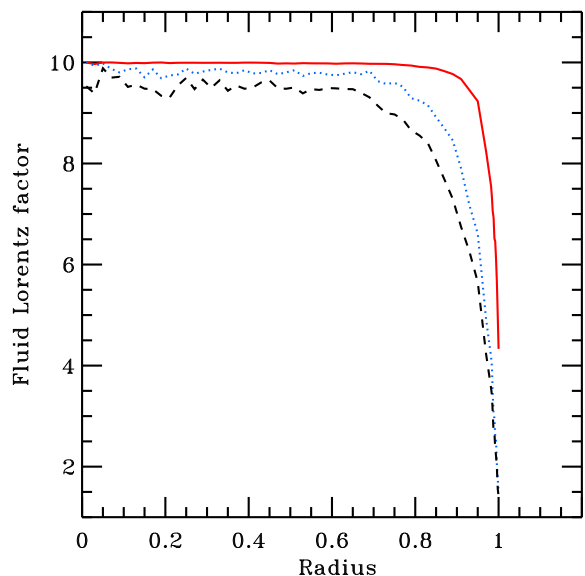,width=7.8cm}}
\caption{Fluid Lorentz factor versus cylindrical radius of the jet $r$ for  Example 1
at various times $t = 6$ (upper curve),
$t = 10$ (middle curve), $t = 20$ (lower curve).
}
\label{fig:lorentz}
\end{figure}

 The total  (cumulative) energy release as a function of time is shown in
 Fig.~\ref{fig:evol}.
 We observe a reasonably fast breeding with e-folding time
$t_{\rm e}  \sim 0.6$. The active layer is rather thin:
a half of the energy release is concentrated within $\delta r\sim 0.02$ from the
jet boundary. At $t \sim 6$ the regime changes: the external
shell decelerates (see Fig.~\ref{fig:lorentz}) and the active layer gets wider
($\delta r  = 0.05$ at $t = 8$ and $\delta  r   = 0.08$ at $t=20$).
The cascade breeding slows down as the photon path length through the
cycle   increases. The total energy release into photons reaches
19 per cent of the total jet energy at the end of simulation at $t = 20$.
Simulation demonstrated a significant pair loading at the late stage:
the Thomson optical depth of produced pairs across the jet has reached $\sim 3
\ 10^{-5}$ which exceeds the initial depth of electrons associated
with protons by 25 per cent  (assuming proton dominated jet, see Eq. [\ref{eq:taut}]). 
Most of these pairs are loaded in the outer region of the jet ($r > 0.8$), where 
they dominate the number density of original electrons by a factor of 8.
 
% We cannot follow further evolution because of neglecting hydrodynamical effects.

%%%%%%%%%%%%%%%%%%%%%%%%%%%%%%%%%%%%%%%%%%%%%
%%%%%%%%%%%%%%%%%%%%%%%%%%%%%%%%%%%%%%%%%%%%%

\begin{figure}
\centerline{\epsfig{file=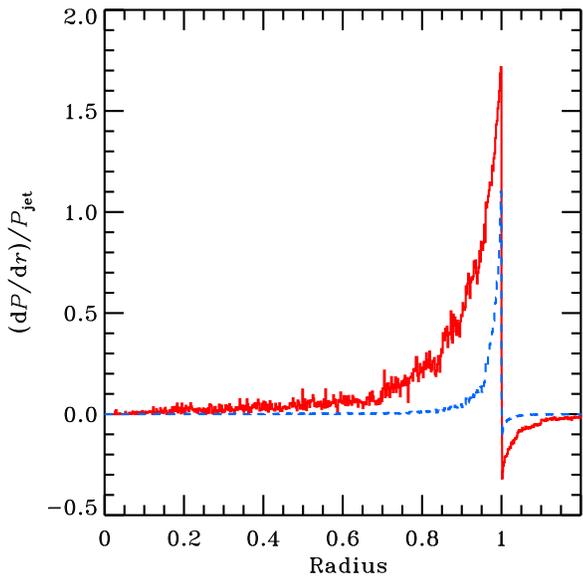,width=7.8cm}}
\caption{Momentum exchange between the fluid  and photons for  Example 1.
A narrow distribution (dashes) corresponds to time $t=6$, and a wider distribution
is for $t=20$.
}
\label{fig:exchange}
\end{figure}

 While the jet decelerates in our model, the external environment is fixed
at rest. In reality, it undergoes a radiative acceleration. Fig.~\ref{fig:exchange} shows
the momentum exchange between matter and photons. The momentum
transferred to the external environment is an order of magnitude less than
momentum losses of the jet to radiation. The momentum gained by
the external shell $\delta r  = 2 \  10^{-3}$ around the jet boundary is
$\Delta P \approx  10^{48}$ erg/$c$.  The volume of the
innermost external shell
is $2\pi  \Rj \ \delta r \ \Delta z \approx 10^{47}$ cm$^{3}$.
Therefore the volume density of the
transferred momentum is $\sim 10$ erg\ cm$^{-3}$/$c$.
If the density of external medium  exceeds $10^4$ cm$^{-3}$
(i.e. the energy density $n \mp c^2$ is higher than $10$ erg\ cm$^{-3}$),
we   can neglect the effects of its acceleration
by the deposited momentum.

\begin{figure}
\centerline{\epsfig{file=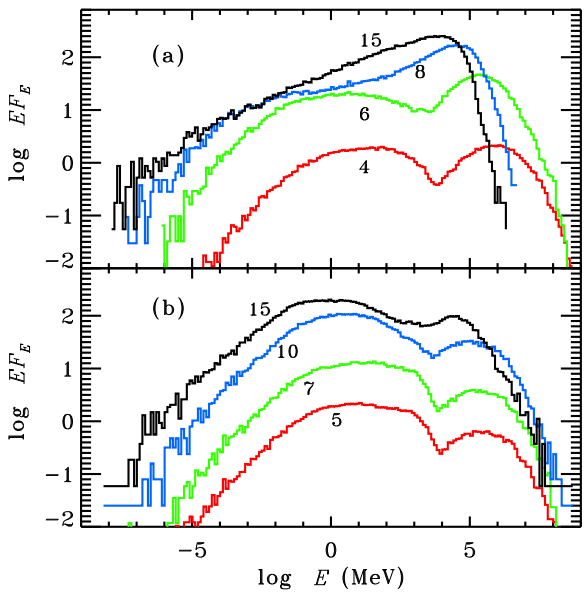,width=7.8cm}}
\caption{Evolution of the instantaneous (jet frame)
photon spectra within the emitting medium. 
(a) Example 1, (b) Example 2.
The observed spectra should be close to the latest ones.
Time in units $\Rj/c$ is marked next to the corresponding curves.
}
\label{fig:spectra}
\end{figure}

The evolution of the photon instant spectrum (i.e. the spectrum of photons
which are in the volume at the moment) is shown in Fig.~\ref{fig:spectra}a.
Early spectrum
demonstrate two distinct components:  the TeV Comptonization peak (mainly
Compton scattered external isotropic photons) and the synchrotron maximum.
After $t \sim 6$ the spectrum changes: the main peak moves to lower energies
and the synchrotron peak declines. The reason for such evolution is evident:
the system enters a nonlinear stage, because the synchrotron radiation of
the cascade exceeds the initial soft photon field. Comptonization losses
increase, while the synchrotron losses do not change.

Note, that an observer cannot see the high-energy part of early spectra
because the isotropic radiation field $\Fiso$ is opaque
for  photons with energy $\phe > 1/\Theta$. The observed
spectrum should have such a cutoff in the late spectra.  
A detection of a sharp cutoff at tens of GeV 
would be the evidence for gamma-rays origin in quasars (blazars) 
at the scale of the broad emission line region.

%%%%%%%%%%%%%%%%%%%%%%%%%%%%%%%%%%%%%%%%%%%%%
%%%%%%%%%%%%%%%%%%%%%%%%%%%%%%%%%%%%%%%%%%%%%

%%%%%%%%%%%%%%%%%%%%%%%%%%%%%%%%%%%%%%%%%%%%%
%%%%%%%%%%%%%%%%%%%%%%%%%%%%%%%%%%%%%%%%%%%%%

\subsection{Example 2. Strong magnetic field}
\label{sec:example2}

We consider the same parameters as in Example 1, but take $\etab=1$
implying magnetically dominated jet with Poynting flux 
$\Lb = 10^{45} \ \mbox{erg\ s}^{-1}$ (for a two-sided jet). 
The comoving value of magnetic field is 3.5 G.
The ratio of Compton to synchrotron losses at such parameters is low 
(see Section \ref{sec:step3}). Therefore, the system remains sub-critical
and the simulations have shown no supercritical photon breeding at such conditions.
Our analytical calculations show that the maximum of $A$  is about 0.5. 
The situation changes if we take a softer spectrum of the 
isotropic component with $\alphaiso=1$, then $\max A=3.5$. 
If we instead increase  the soft photon density to $\etaiso=0.2$, then $\max A=1.2$.  
However, the easiest way to achieve criticality is to increase  the jet Lorentz factor
to $\Gamma=20$, which gives  $\max A =8.4$.  All this options increase 
the importance of Compton cooling relative to the synchrotron (see eq.[\ref{eq:urub}, \ref{eq:c3app}]).

In the numerical simulations, we followed the last alternative.
In this case, the comoving magnetic field is 1.75 G. 
%However, at $\Gamma = 20$ (and comoving magnetic field 1.7 G 
%corresponding to $\etab=1$) the supercritical regime does occur,
%because the energy density of the external photons in the jet frame 
%grows as $\Gamma^2$ and Compton cooling becomes more important.
The simulations have shown that the active layer is wider
than in Example 1: $\delta r \sim 0.05$ at
the beginning of the evolution (therefore this case is more stable
against the mixing of the boundary layer).
The exponential growth is slower  (see Fig.~\ref{fig:evol}), 
but the final energy release is the same as in the previous case.
The pair loading is an order of magnitude smaller than in Example 1.

The hard to soft evolution of high-energy peak of the photon spectrum
shown in Fig.~\ref{fig:spectra}b
spans almost all range of peak energies observed in blazars.
The latest spectrum peaks in MeV range as in MeV blazars.

\section{Discussion}
\label{sec:disc}

We have demonstrated that a supercritical runaway cascade   develops under
reasonable conditions and can convert at least $\sim$ 20 per cent of the jet kinetic 
energy into radiation. This is certainly not an
ultimate value: with our simplified model we are able to reproduce only
the initial stage of the evolution.
Indeed, our cylindric shells decelerate as a whole with constant Lorentz
factor along $z$ axis. Therefore, once the outer shells decelerate, the
cascade breeding slows down everywhere. Actually, the radial gradient
of $\Gamma$ should depend on $z$ and a slow breeding at larger $z$  can
coexist with a fast breeding at a smaller $z$. 

The behaviour of the jet  is strongly non-linear because the process is 
very sensitive  to a number of details including geometry of the magnetic field,
density of the external environment, density of the isotropic soft  photon field, etc. 
The effect of inhomogeneities, which are exponentially amplified,  can be dramatic, 
particularly taking a form of flares and moving bright blobs and resulting in 
formation of internal shocks.
A more realistic model should include a detailed treatment of fluid
hydrodynamics (at least in 2D) coupled with the electromagnetic cascade.

The model can reproduce the high-energy component of blazar radiation. 
On the other hand,
examples presented in this work do  not reproduce the low-energy
synchrotron components as prominent as observed in blazars.
It is possible that the synchrotron bump is produced at much larger 
distances from the central source than  the Compton component.
%The general impression is that our simulated spectra are qualitatively
%similar, but flatter and smoother than observed.

A certain problem can appear if the jet boundary is turbulent \citep[see e.g.][]{a99}: 
the thin active zone at the jet boundary layer as in above examples does not exist in this case. 
Then one still can expect to obtain the supercriticality at a moderate
transversal opacity. However the breeding cycle in this case would take longer than
$\Rj/c$. If the electromagnetic cascade grows only along the jet, the 
active range  of $z$ could be insufficient to provide the growth by orders of 
magnitude. An issue to be studied is whether the cascade can grow {\it with time }
at a fixed $z$. This could be due to a spatial feedback at step 5: a photon 
from the external environment moves upstream to a smaller $z$ than the point 
where a parent  photon has been produced at step 3. It is clear that such 
feedback is weak, but in the case of a jet the time is unlimited.
Note that even if the supercriticality is not reached, the process can 
amplify in subcritical regime the high-energy output of charged particle acceleration.

Now let us try to characterize in general terms the mechanism we are dealing with.
First of all, it belongs to a class of supercritical runaway phenomena
like neutron breeding in a nuclear pile or a nuclear explosion. Such phenomena
still seems rather exotic in astrophysics (let alone ``trivial'' nuclear
explosions of supernovae or of the accreting matter at the surface of neutron
stars). To our knowledge there exist only a few works considering
such kind of phenomena. \citet{ss91} discovered
with numerical simulations a supercritical
behavior of electromagnetic cascade in a cloud of ultra-relativistic protons
with sufficient compactness. In that case the energy is stored in non-radiating
protons and the super-criticality appears in the energy transfer from protons to
pairs and photons through photo-meson production.
Later this mechanism was confirmed analytically by
\citet{km92}. However, at that time, there was no clear
astrophysical situation providing   proper conditions. Recently \citet{k02}
have found that a possible site for such phenomenon could be
a highly relativistic shock in gamma-ray bursts.

In this work  \citep[see also][]{s03} we propose a different kind of
a supercritical process where the energy is extracted by particles directly
from the kinetic energy of the fluid. In principle, the protons can also  participate
in such mechanism, especially, if a scheme of \citet{d03} with
$p + \gamma \rightarrow n + \pi^+$  charge exchange works at given conditions. In this
case we would have a unified mechanism, where the electromagnetic cascade is fed
by the fluid bulk motion directly and through the high-energy nucleons.

In any case, supercritical models of energy conversion look promising for
explanation of such violent phenomena as blazars and gamma-ray bursts
because the supercriticality does produce violent effects.

\section{Conclusions}
\label{sec:concl}

We have proposed and studied a novel photon breeding mechanism   of the high-energy 
emission from relativistic jets.  We showed  that a relativistic jet moving through 
the sufficiently dense soft radiation field inevitably undergoes transformation into a luminous state. 
We have considered the application of this mechanism for the 
AGN jets, while actually it may also work in microquasars and gamma-ray bursts,
if the latter are associated with the well formed narrow  jets.
In general, the mechanism can be characterized as a viscous dissipation of the 
kinetic energy of the jet into high-energy photons. 
We showed that at least 20 per cent of the jet energy can be converted  
into high-energy radiation.
From the dynamical point of view the mechanism is a supercritical 
process, which is very similar to the chain reaction in the supercritical 
nuclear pile. The subject to exponential breeding in our case is electromagnetic cascade, 
particularly high-energy photons which create a viscosity between the jet and the external 
environment. 

%There is a concern, however, if there is enough time for the cascade to fully develop. 
In the case, when the exponential photon breeding does not occur, 
the process still could act as a subcritical amplifier for the high-energy output of 
internal shocks or other mechanisms of particle acceleration. 
Alternatively, it can produce a population of hot  electrons/positrons  which 
then can participate in the Fermi I type of acceleration of charged particles. 

The photon breeding mechanism  works very efficiently at the following conditions:
\begin{enumerate}
\item
Jet Lorentz factor exceeding 3--4.
\item 
Weak magnetic field and/or large 
density of the external soft isotropic radiation field. 
\item
If the magnetic field is strong (equipartition or magnetically dominated 
jet), the supercritical regime requires a dense soft radiation component 
(in the jet frame), which requires either a large fraction of scattered photons or a
large jet Lorentz factor.
Alternatively, an additional source of soft photons, e.g.  the synchrotron radiation 
from other forms of the jet high-energy activity, e.g. by particle acceleration in internal shocks,
is needed. The required intensity of this activity is much 
smaller than the final high-energy output.
\item
The presence of a sharp boundary of the jet, with the depth of the transition layer 
less than a few per cent of the jet radius (which is still several orders
of magnitude larger than required for charged particle acceleration). 
If the jet does not have a sharp boundary, then the formation of a supercritical 
regime is still possible at large distances, where the photon-photon pair production 
opacity declines.  
\end{enumerate}

%----------------------------------
%\section{Acknowledgments}
\section*{Acknowledgments}
 
The work is supported by the 
Russian Foundation for Basic Research grant 04-02-16987,
the Jenny and Antti Wihuri Foundation, the
Vilho, Yrj\"o and Kalle V\"ais\"al\"a Foundation, and  
the Academy of Finland grants  107943 and 102181.
%----------------------------------

%----------------------------------
\label{lastpage}

\end{document}